\documentclass[a4paper,12pt]{article}
\usepackage[margin=2.5cm,letterpaper]{geometry}
\usepackage{authblk}
\usepackage{setspace}
\usepackage{subfig}
\usepackage{amsmath}
\usepackage{amsfonts}
\usepackage{amssymb}
\usepackage{caption}
\usepackage{graphicx}

\usepackage[super,comma,sort&compress]{natbib} 
\usepackage[english]{babel}
\usepackage[T1]{fontenc} 

\title{The effect of gravity on the stability of an evaporating dichloromethane liquid film}

\author[1]{Aneet D. Narendranath}
\author[2]{James C. Hermanson}
\author[3]{Robert W. Kolkka}
\author[3]{Allan A. Struthers}
\author[1]{Jeffrey S. Allen \thanks{jstallen@mtu.edu}}
\affil[1]{Mechanical Engineering-Engineering Mechanics, Michigan Technological University, Houghton, MI}
\affil[2]{Aeronautics \& Astronautics, University of Washington, Seattle, WA}
\affil[3]{Mathematical Sciences, Michigan Technological University, Houghton, MI}

\begin{document}
\maketitle 

\begin{abstract}
\noindent Zero gravity evaporation of a Dicholoromethane (DCM) liquid film is explored. The resulting film dynamics are presented and a criterion for stable films is described based on the long wave theory. It is concluded that films subject to long wave instabilities shows the appearance of the mode of maximum growth rate at rupture, irrespective of the initial condition or domain size conditions. Films stable in Earth's gravity are destabilized in zero gravity.

\end{abstract}

\section{Impact of liquid films}
Liquid films and coatings find applications in several areas of technology, physics and biology as their flows occur over a large range of length scales \citep{Craster2009a}. Coatings receive widespread attention in several industries. They are used as lubricants, paint for rust prevention and aesthetics in architecture, in time release capsules and tablets in the medical industry, the paper and pulp industry and the microelectronics fabrication industry. Liquid films find applications in the food processing industry \citep{Pehlivan2012a, Moresi1988a}, cooling of server towers (which was a \$36 Billion industry in 2007) \citep{Agostini2007a, Darabi2003a, Marcinichen2010a} and for improved oil recovery from petroleum reservoirs \citep{Schramm1992a, Bergeron1993a}. A report from the Brookhaven National Laboratory estimated that in excess of \$100 billion per year, is spent on the prevention and remediation of rust damage \citep{Cheremisinoff2003a}.\\

One of the earliest commentaries to study liquid films was by Rayleigh\citep{Rayleigh1916a} wherein he described convective cells in \emph{spermacetti} heated from below and the corresponding wavelength associated with this convection pattern. It was suggested by Rayleigh that this convection was driven by surface tension. This was corrected later, notably, by Pearson \citep{Pearson1958a} who suggested that given the rigid-rigid boundary conditions used by Rayleigh could not have allowed for surface tension gradients to drive the convection. Rayleigh's observations were deemed to be due to buoyancy effects. Pearson conducted an examination of liquid films heated from below and subject to rigid-free boundary conditions at the bottom and top respectively to uncover surface tension driven convection. These convection cells were thereafter referred to as Pearon cells. Scriven and Sterling provided numerical methodology for a similar situation as that which gave rise to the ``Pearson cells'.  In their case, the surface was assumed to be infinitesimally deformed via long waves. Scriven and Sterling's work paved the way for the long wave theory. The long wave theory (LWT) was further explored shortly after by Benney \citep{Benny1966a, Benny1966b} to describe the interactions of long wavelength disturbances in falling films.\\

The study of long wave interactions on the surface of liquid films gained significant impetus when Davis et al. \citep{WandD} and later Burelbach et al.\citep{Burelbach} developed a non-linear evolution equation to describe the film dynamics of an evaporating liquid film. The essense of the Navier-Stokes' equations was rolled into this evolution equation. This equation has been used since for a variety of problems concerning liquid films that could be categorized under problems of the lubrication approximation type/long wavelength phenomena.\\

This theoretical treatment can be traced far back to Benney \citep{Benny1966b}. The evolution equation, as a result of applying the LWT to the Navier Stokes' equations has been successfully used to analyse liquid film dynamics and several instances can be found in literature \citep{Krishnamoorthy1995a, Vanhook1997a, Oron1997a, Oron2000b}.\\

The general trend in literature since the development of the evolution equation has been the simulation of non-evaporating liquid films via LWT/evolution equation \citep{Krishnamoorthy1995a, Vanhook1997a, Oron2000b, Sultan2005a, Dietzel2009a}. The stability of evaporating liquid films were analysed in two-dimensional space by Burelbach \cite{Burelbach} and in three-dimensional space by Oron \citep{Oron2000b, Oron2000c}. The former lacked three-dimensional simulations while the latter did not attempt a solution to the complete evolution equation for evaporating liquid films and neglected  mass loss and destabilization by vapor recoil in favor of disjoining pressures.\\

There is a paucity of research on zero gravity film dynamics. Pradhan and Samal \citep{Pradhan1987a} and Straughan \citep{Straughan1989a} performed some normal mode analysis on the full Navier Stokes' equations in zero gravity.  Studying zero gravity film dynamics is important as zero gravity environments help isolate a system from terrestrial gravity effects. Understanding the instabilities that affect fluids and liquid films in zero gravity is important as capillary forces could become significant and could govern the behavior of the liquid system. For instance, propulsion systems or water recovery systems that employ or recover liquids would behave differently in low gravity as compared to an Earth like gravity field. Evaporation in such systems may not allow the liquid and the gas phases to separate into two distinct components due to a lack of buoyancy \citep{Ostrach1982a}.\\

In this article, a modified evolution equation describing evaporating liquid film dynamics is solved numerically as an initial value problem to describe a slowly evaporating Dicholoromethane liquid film in zero gravity and Earth's gravity. The importance of gravity in dictating film dynamics is studied. \\

\section{Evolution equation describing dynamics of an evaporating liquid film}

The evolution equation derived from the LWT, describing an evaporating Newtonian liquid film is given by equation \ref{evolution1}. The non-dimensional parameters, viz., $E, S, Ga, E, D, K, M, Pr, Ra$  are expressed in table \ref{tab:nomenclature}.

\begin{align} \label{evolution1}
 h_T + S \nabla \cdot (h^3 \nabla \nabla^2 h) - \frac{Ga}{3} \nabla \cdot (h^3 \nabla h)  \nonumber \\
+ \nabla \cdot \left[ \left( E^2 D^{-1} \frac{h^3}{(h + K)^3} + KM\text{Pr}^{-1} \frac{h^2}{(h + K)^2} \right) \nabla h \right] \nonumber \\
 \frac{5 Ra}{48 \text{Pr}} \left[ \frac{K^2}{(h + K)^2}h^4 + h^4\right] \nabla h = 0 
\end{align}

Using $S$ to scale the various terms in the evolution equation and using the linear stability theory, the wavenumber corresponding to the fastest growing wavenumber is derived and presented in equation \ref{wavenumber_eqn}.  The fastest growing wavelength is merely $2\pi/q_\text{max}$ where $2\pi$ is the length of the domain.



\begin{align} \label{wavenumber_eqn}
 \text{q}_\text{max} = \sqrt{\frac{-Bo}{2} + \frac{\delta}{(h + K)^3} + \frac{m}{2 h (h + K)^2}} 
\end{align}

Here $Bo = G/S, \delta = E^2/DS$, $m = MK/\text{Pr}S$ and $R = 5 Ra/48\text{Pr}S$ are respectively the Bond number, evaporative effects of mass loss and vapor recoil, thermocapillarity number. 

There are several advantages to scaling the evolution equation with the surface tension number, $S$. The evolution equation is generally scaled with a combination of $Ga$ and $S$. However, involving the Galileo number, $Ga$, in the scaling leads to undesirable divide by zero errors when exploring film dynamics in zero gravity environments. The scaling method used here to arrive at the fastest growing wavelength allows for zero gravity film dynamics to be explored. 

\subsection{Validation}



The evolution equation has been verified and validated against literature \citep{Krishnamoorthy1995a, Oron2000b} by comparing film dynamics for non-evaporating cases. We use slowly evaporating films ($E=0.0001$) whose fastest growing wavelength are not far removed from non-evaporating liquid films. The fastest growing wavelength for various non-evaporating, evaporating situations is tabulated in table \ref{tab1}. This serves as a comparison with literature \citep{Krishnamoorthy1995a, Oron2000b} and validation of our expression (termed ``MTU/UW'') for the fastest growing wavelength as in equation \ref{wavenumber_eqn}.

Table \ref{tab1} shows a comparison with the fastest growing wavelength as calculated by Oron \citep{Oron2000b}.  The first row is a validation of our expression for the fastest growing wavelength, equation \ref{wavenumber_eqn}. The second and third rows of table \ref{tab1} show the influence of increasing the number of mechanisms (viz., lack of gravity, evaporative mass flux, non-equilibrium at the interface) that affect the dynamics of a liquid film.

It is observed from table \ref{tab1} that for a slowly evaporating liquid film with an $E=0.0001$, the fastest growing wavelength is not affected strongly by evaporative mass flux and vapor recoil when the $Ga=0.333$. However, it is noticed from table \ref{tab1} that in the case of zero gravity evaporation, the film is susceptible to shorter wavelengths for slow evaporation rates.

The evolution equation \ref{evolution1}, was solved as an initial value problem with periodic boundary conditions with the LSODE solver \citep{Hindmarsh1987a} in the Mathematica environment \citep{Struthers2010a}. We compared our numerical results with those of Oron \citep{Oron2000b} and  Krishnamoorthy\citep{Krishnamoorthy1995a} and have concluded that our code produces results within $5\%$ of literature. We used film profile plots in conjunction with discrete Fourier transform plots (with their DC component zeroed) to analyse the film structure and the frequency trends of evaporating DCM liquid films.


\section{Definition of rupture}
It has been propounded that when a film ruptures on a substrate, an adsorbed layer is left behind whose depletion is a function of the wetting nature\citep{Oron2001a} of the film on this substrate. Rupture is defined as that juncture in time when the the film recedes from the surface depending on the surface wetting characteristics. Disjoining pressure terms would need to be included in the evolution equation, \ref{evolution1} to study film rupture via dewetting on the surface.

We do not include the effect of wetting via Van der Waal's forces in the evolution equation. From hereon, we define rupture as that juncture in the computation when the effect of various non-linear terms in the evolution equation is to induce numerical stiffness at small film thicknesses. In our simulation cases, rupture occurs when the film thickness reaches about 3-4\% of the initial film thickness. Physically, this rupture thickness translates to about $250 \mu m$ for the dichloromethane film we simulate. From the discrete Fourier transform (DFT) plots (eg. figure \ref{fig_ic_effect}), at rupture, the film profile shows the appearance of high frequency Fourier modes. The utility of the DFT plots is enabling the realization that as the film progresses toward rupture, higher frequencies appear. The fastest growing wavelength is the dominant mode. A cascade of higher frequencies is seen as gray spots. The DFT plots show all those frequencies that account for $10\%$ or more of the total energy at that stage in time.

\section{Dichloromethane liquid film evaporating in zero gravity}

\begin{figure} 
   \centering
    \subfloat[]{\includegraphics[width=0.45\linewidth]{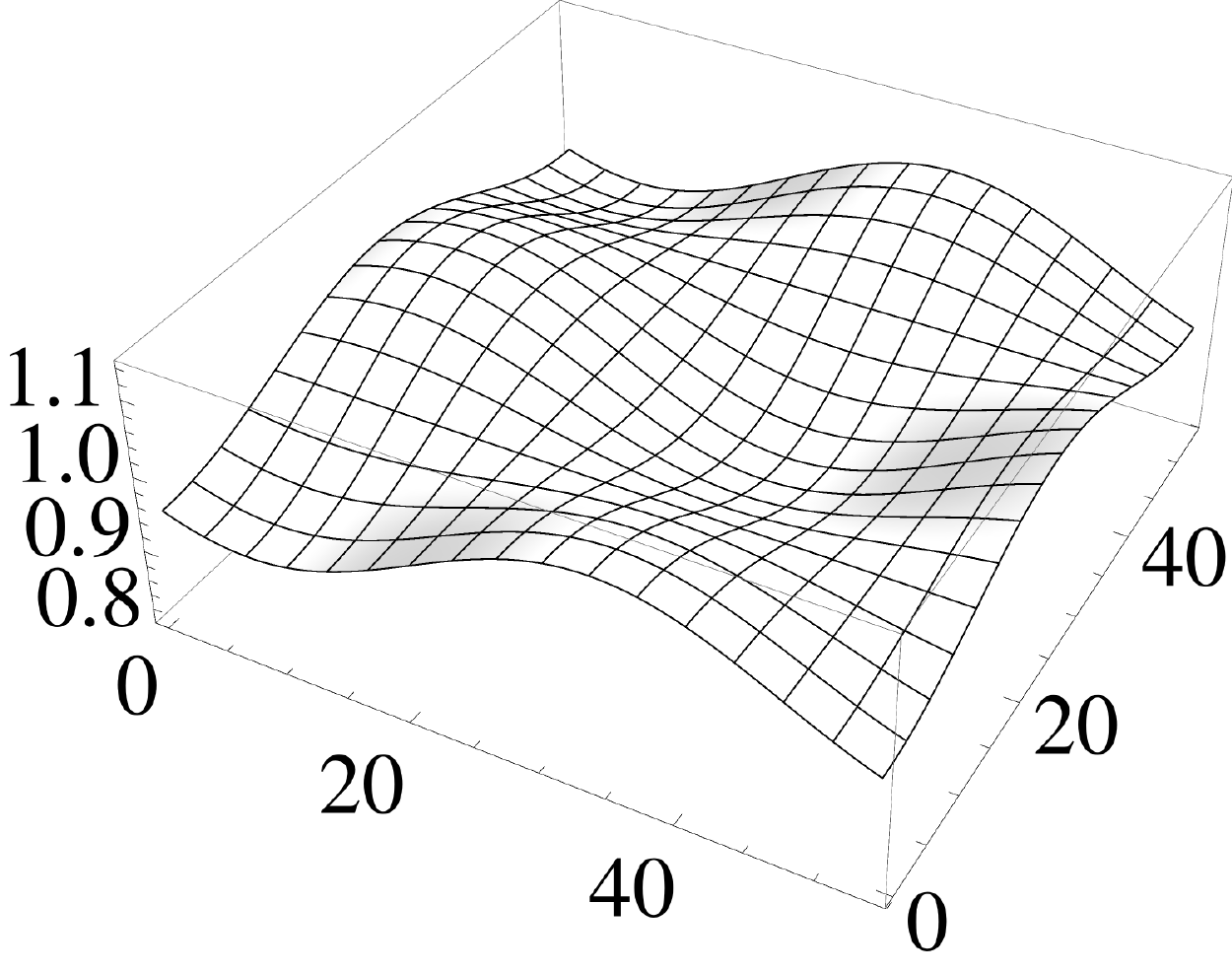}\label{fig:1a}} 
    \subfloat[]{\includegraphics[width=0.45\linewidth]{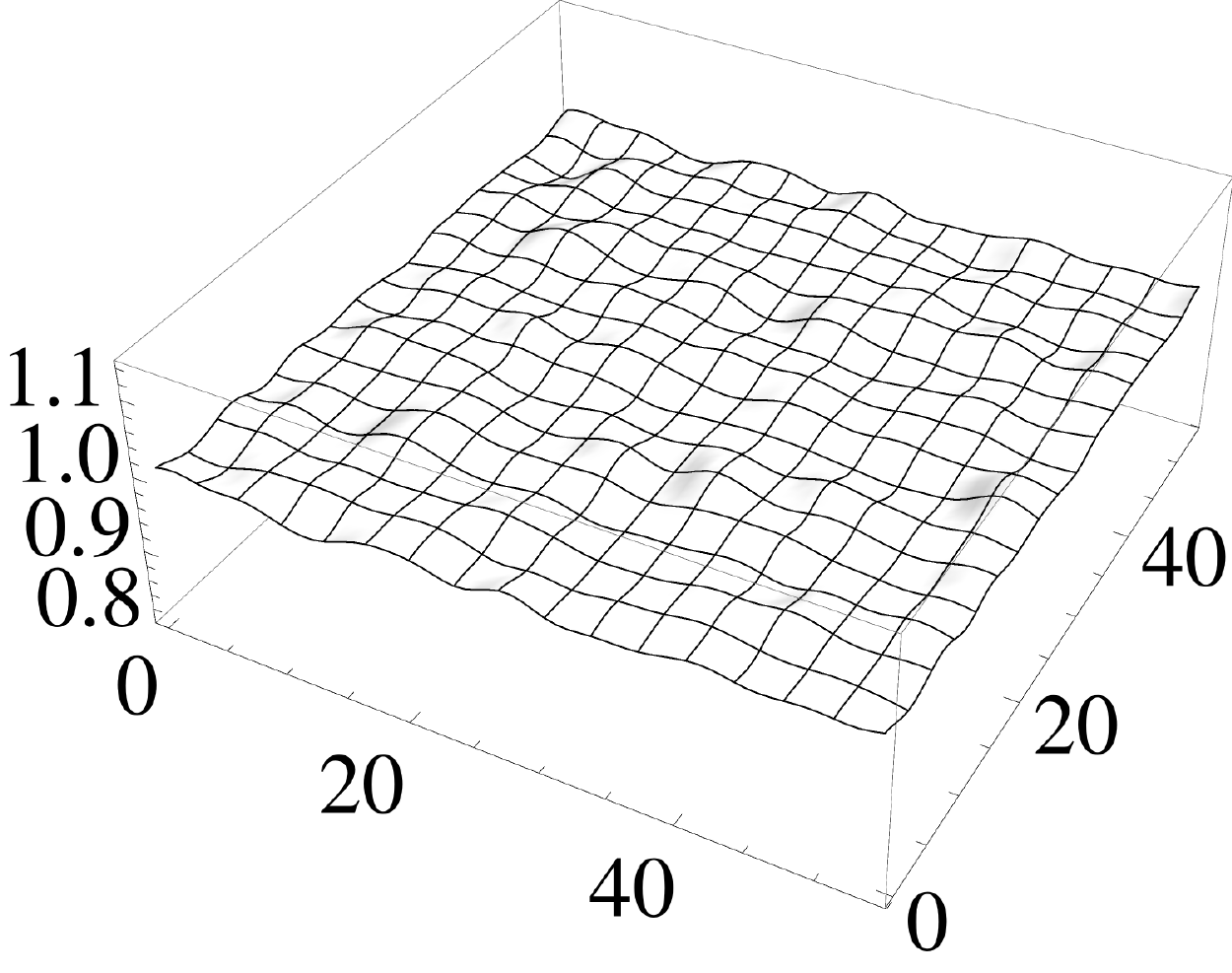}\label{fig:1b}} \\
    \subfloat[]{\includegraphics[width=0.45\linewidth]{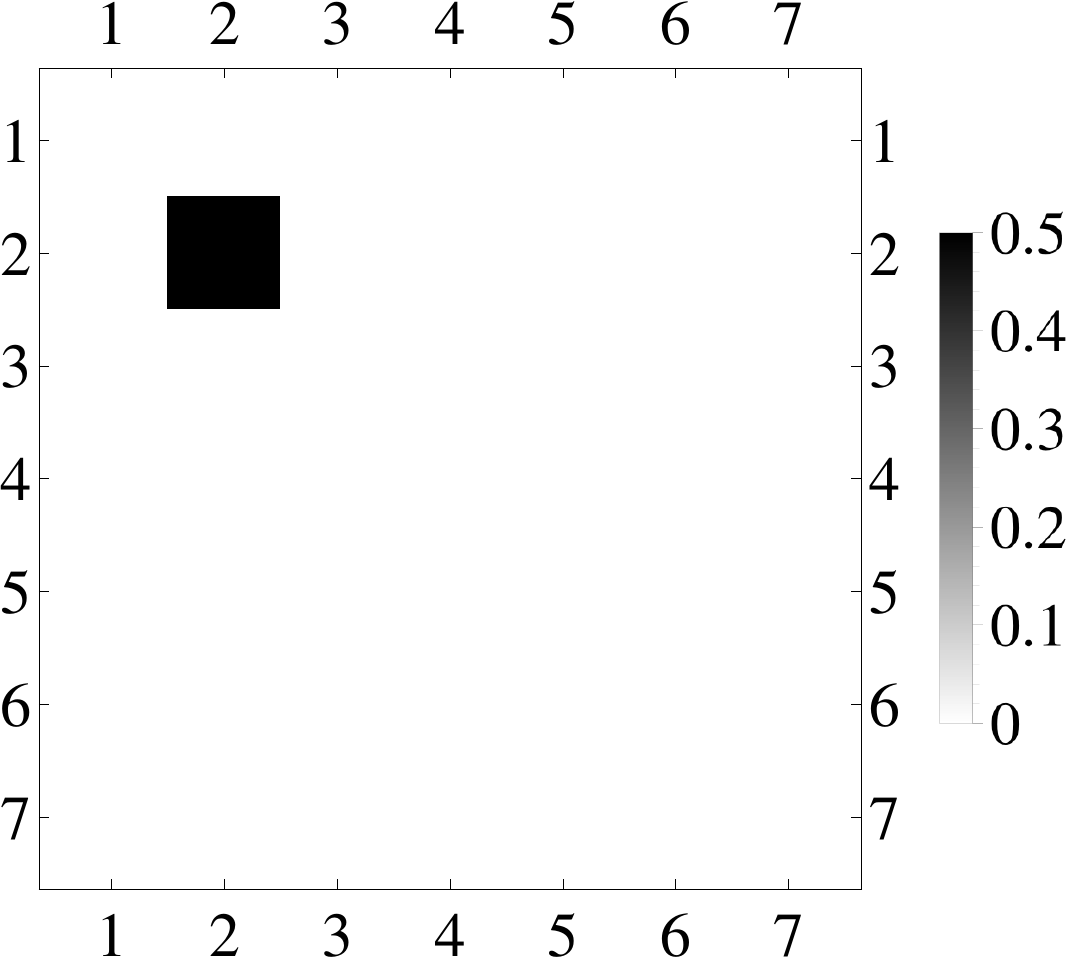}\label{fig:1c}}
    \subfloat[]{\includegraphics[width=0.45\linewidth]{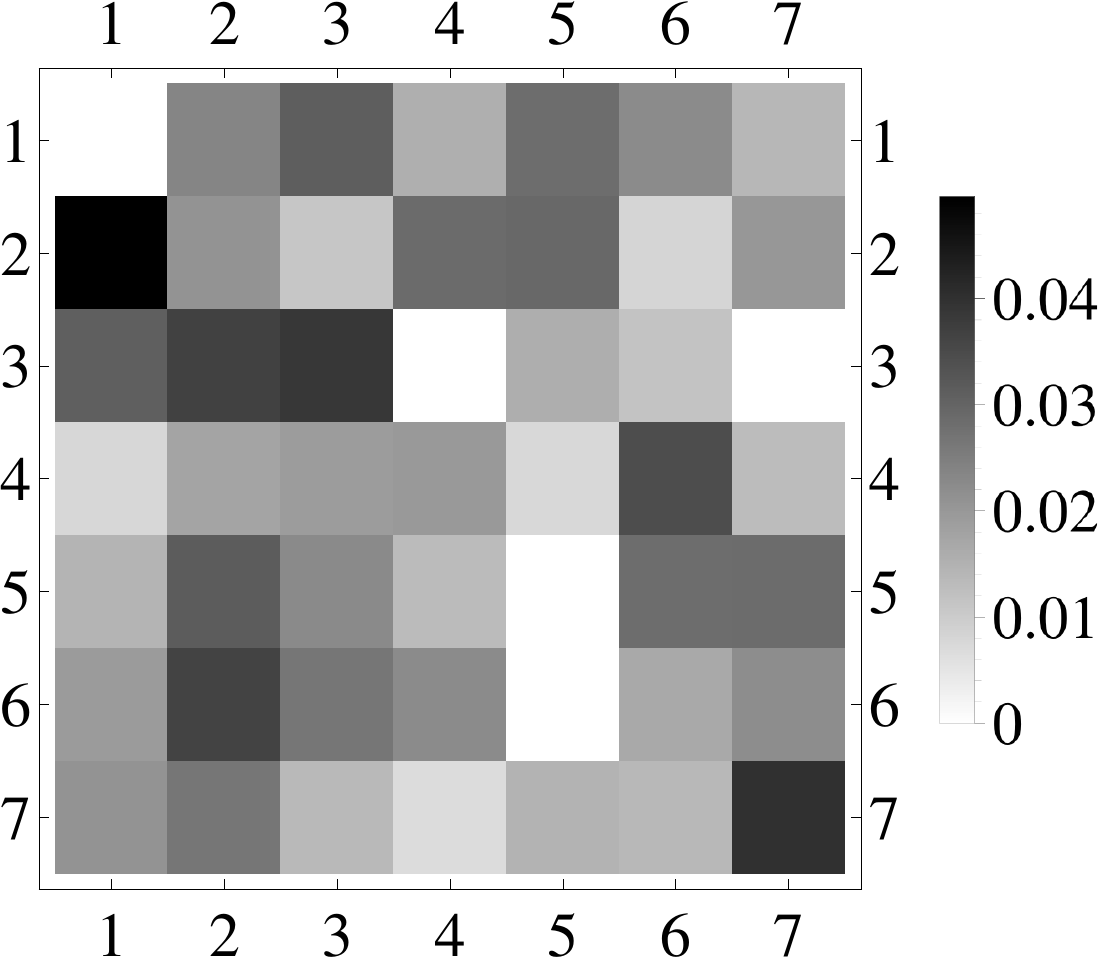}\label{fig:1d}}
   \caption{Two different initial conditions and their discrete Fourier transform frequency distributions. In fig \ref{fig_ic}\subref{fig:1a} a cosine initial condition $1 - 0.05 \left[ \cos [2 \pi x/L] + \sin [2 \pi x/]\right] \cos[2 \pi y/L]$ is used to excite a quiescent film. In fig \ref{fig_ic} \subref{fig:1b} a uniform random perturbation is used to excite the quiescent film.}
   \label{fig_ic}
  \end{figure}

Two different initial conditions are used to perturb a DCM film evaporating in zero gravity. Figure \ref{fig_ic} a smooth cosine initial condition and it's discrete Fourier transform (DFT) frequency distribution is compared against a uniform random distribution. The smooth initial condition is generally used in literature.

 The evolution equation \ref{evolution1} is solved to simulate a 2.35mm thick Dichloromethane liquid film which is slowly evaporating. The non dimensional parameters for the Dichloromethane liquid film are compared with that of the mathematical fluid in table \ref{dcm_math_comparison}. The film is placed in a square domain whose side length is equal to about $5$ inches ($\lambda_\text{max} \approx 5$ inches for a $2.35$ mm DCM film). When the film is perturbed by a smooth initial condition or a uniform random initial condition the film ruptures via long wave type/thermocapillary driven structures. The end result in figure \ref{fig_ic} shows a cascade of Fourier modes in zero gravity. The dominant mode in the DFT plot is that which corresponds to the fastest growing wavelength, with the maximum growth rate.

However, when the DCM evaporates in Earth's gravity, the film is resistant to long wave type instabilities as seen by the blank DFT plot in figure \ref{fig_ic_effect} \subref{fig:2a}. Hence a DCM film that is resistant to thermocapillary rupture in Earth's gravitational field is destabilized in zero gravity and is driven to rupture. Another interesting observation is that in Earth's gravity, the complete depletion of the liquid film takes place in about 1 minute. However, in a zero gravity environment, the film ruptures in around 4 seconds.

\begin{figure} 
   \centering
    \subfloat[]{\includegraphics[width=0.45\linewidth]{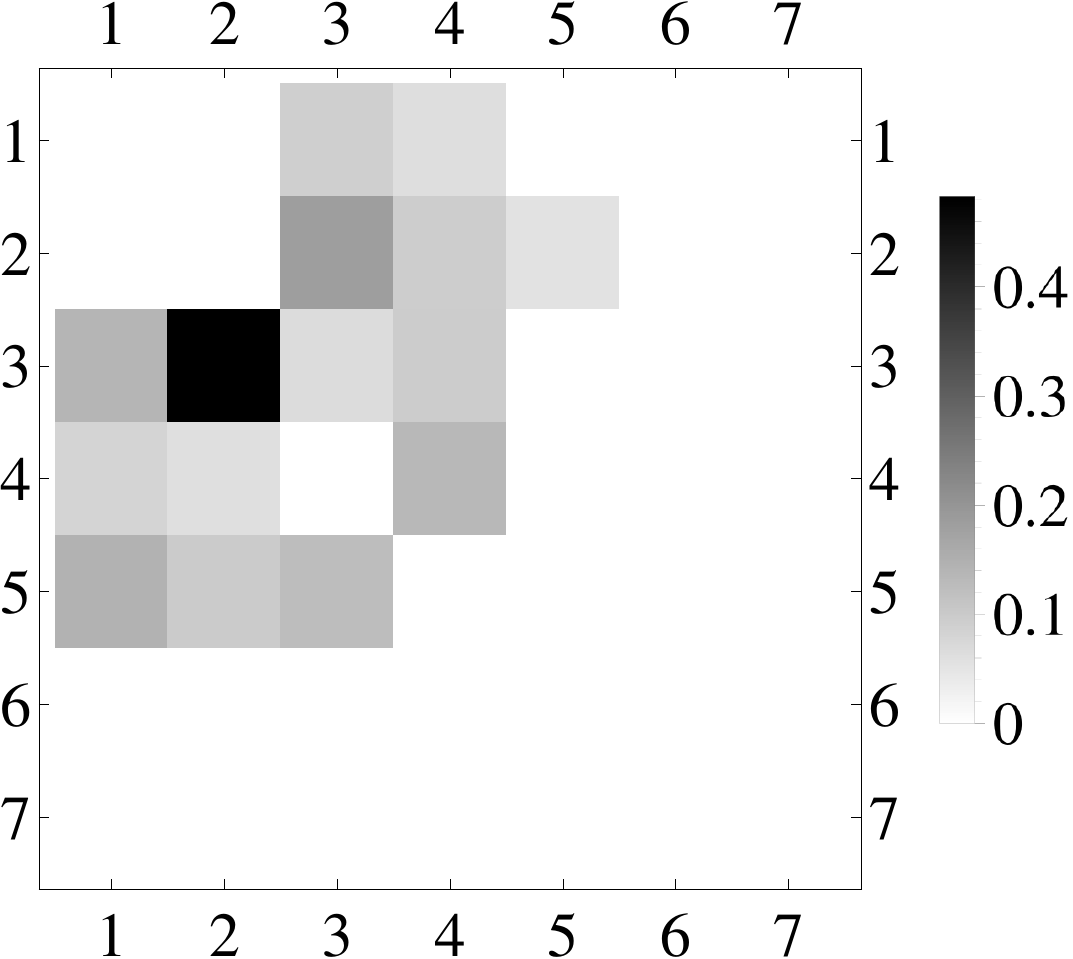}\label{fig:3a}} 
    \subfloat[]{\includegraphics[width=0.45\linewidth]{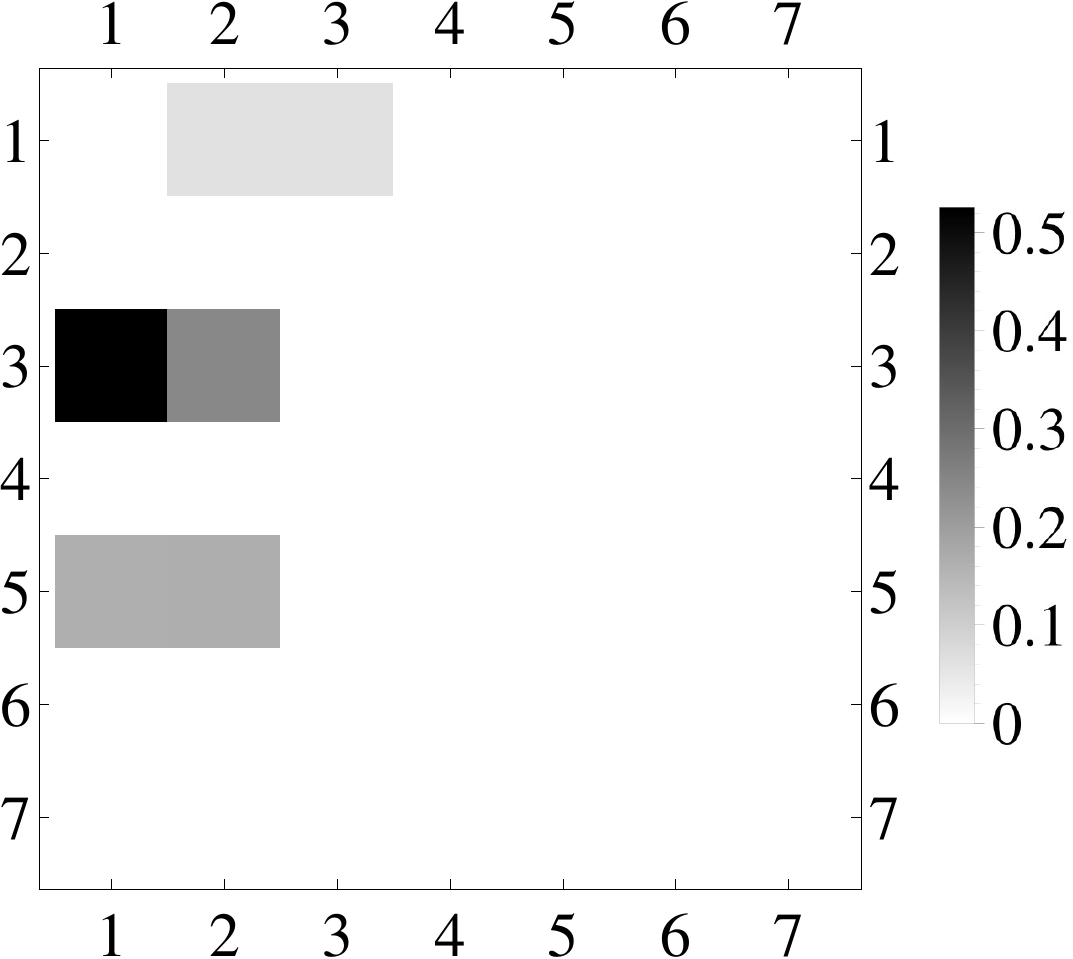}\label{fig:3b}} 
   \caption{The effect of domain size is shown with the cosine initial condition. A square domain is used with $L=2.238\lambda_\text{max}$ in fig \ref{fig_domain_size_effect} \subref{fig:3a} and a rectangular domain is used with $L_x = 2.238 \lambda_\text{max}, L_y=1.641 \lambda_\text{max}$ in figure \ref{fig_domain_size_effect} \subref{fig:3b}. The discrete Fourier transforms show a cascade of frequencies. The dominant mode is that of the fastest growing wavelength.}
   \label{fig_domain_size_effect}
  \end{figure}

The domain size is now changed for this DCM film evaporating in zero gravity. We use a square domain with $L=2.238 \lambda_\text{max}$ and a rectangular domain with $L_x=2.238 \lambda_\text{max}, L_y=1.641\lambda_\text{max}$. When the film is perturbed with a uniform random initial condition a cascade of frequencies with the fastest growing wavelength as the dominant mode emerges as seen in figure \ref{fig_domain_size_effect}.

\section{Criterion for film thickness for long wave destabilization}

Based on gravity stabilization and thermocapillarity destabilization conditions, there exists a criterion for the growth of thermocapillarity driven long wave instabilities. For a certain balance between stabilizing (gravity, surface tension) and destabilizing (thermocapillarity, evaporative mass flux, vapor recoil) forces, the film thickness has an effect on whether or not long wave type instabilities manifest.

The growth rate of long wave instabilities can be used as a metric to define the film thickness criterion for long wave destabilization affecting evaporating liquid films in zero- and micro-gravity environment. The growth rate of long wave instabilities is given by the equation \ref{omega_1}. This is derived from the method of linear stability applied to the evolution equation \ref{evolution1}.

\begin{align} \label{omega_1}
\omega = \frac{\delta  h^3 \text{q}^2}{(h+\text{K})^3}-\text{Bo} h^2 \text{q}^2-h^3 \text{q}^4 +\frac{h^2 m \text{q}^2}{(h+\text{K})^2}+\frac{\epsilon}{(h+\text{K})^2}
\end{align}

\begin{figure} 
   \centering
    \subfloat[]{\includegraphics[width=0.45\linewidth]{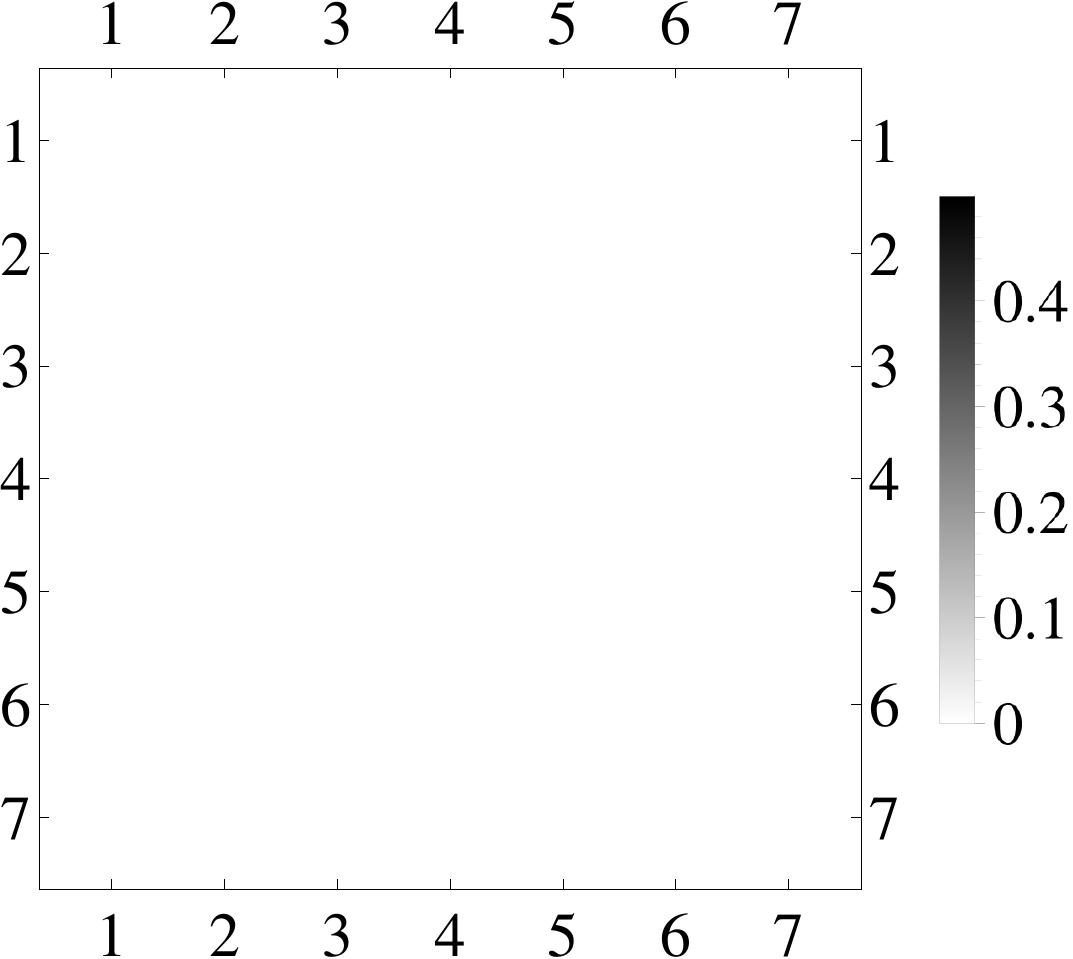}\label{fig:2a}} 
    \subfloat[]{\includegraphics[width=0.45\linewidth]{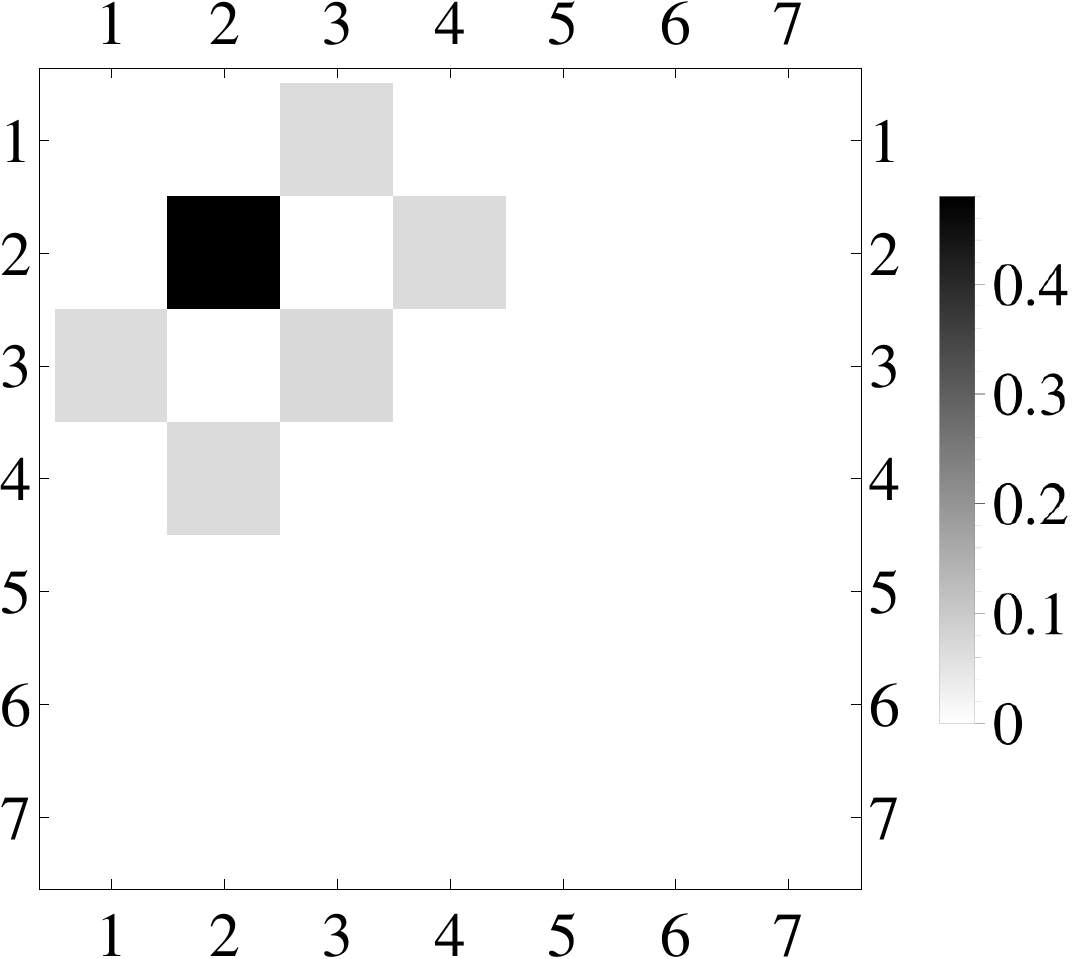}\label{fig:2b}}
   \caption{The effect of gravity on film dynamics for a square domain with $L=\lambda_\text{max}$. In fig \ref{fig_ic_effect}\subref{fig:2b}The discrete Fourier transforms show a cascade of frequencies for zero gravity. The dominant mode is that of the fastest growing wavelength.}
   \label{fig_ic_effect}
  \end{figure}

A positive growth rate ($\omega$) signifies a positive growth rate of long wave instabilities. Positive destabilizing growth rates occur in the case depicted in figure \ref{fig_ic_effect} \subref{fig:2a}. A zero or negative growth rate signifies a damping of long wave instabilities. Negative, damping of instabilities occurs in figure \ref{fig_ic_effect} \subref{fig:2b}.

The growth rate, $\omega$, in equation \ref{omega_1} is function of film parameters $Bo, K, m, \delta, \epsilon$, the excitation wavenumber, q and the mean film thickness $h$. For a given excitation, q, given Bond number, $Bo$ and given strength of thermocapillarity, $m$, the growth rate of long wave instabilities would depend on the mean film thickness, $h$.

\begin{figure}
   \centering
   \includegraphics[width=\linewidth]{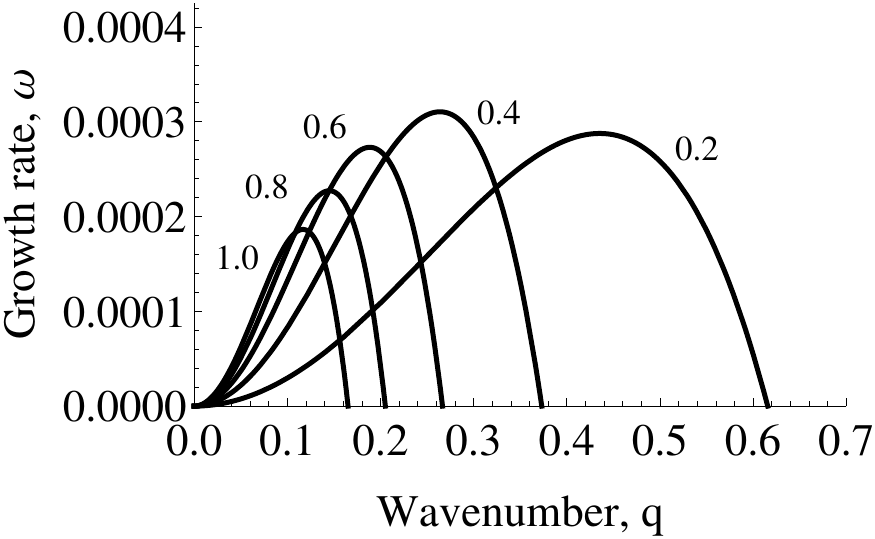} 	
   \caption{The growth rate of long wave instabilities for varying film thickness ($h=0.2$ to $1.0$) for a $2.35$ mm film evaporating in zero gravity ($Bo=0.0$). As the film thickness decreases, the range of perturbation wavenumbers for which long wave instabilities set in increases. Film thickness is scaled with $h_0=2.35$mm.}
   \label{omega_q_1}
  \end{figure}

\begin{figure}
   \centering
   \includegraphics[width=\linewidth]{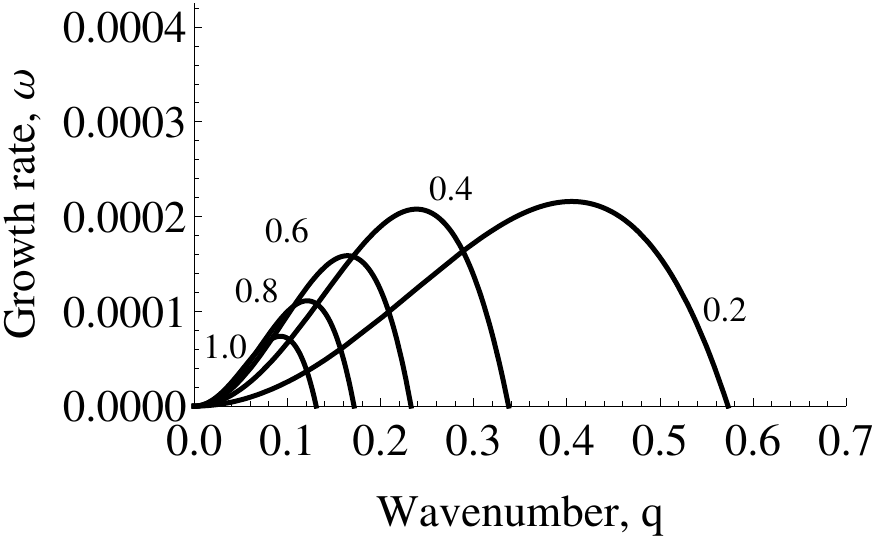}
   \caption{The growth rate of long wave instabilities for varying film thickness for a $2.35$ mm film evaporating in micro-gravity ($Bo=0.01$). As the film thickness decreases, the range of perturbation wavenumbers for which long wave instabilities set in increases. However, as a result of some gravity stabilization, the growth rate of long wave instabilities is lower than in the zero gravity case in figure \ref{omega_q_1}.Film thickness is scaled with $h_0=2.35$mm.}
   \label{omega_q_2}
  \end{figure}

A plot depicting the trend of $\omega$ vs q for different film thicknesses is shown in figure \ref{omega_q_1} for zero gravity with $Bo=0.0$, $m=0.1092$, $\delta=5.19 \times 10^{-7}$ and $\epsilon=6.01 \times 10^{-8}$ and in figure \ref{omega_q_2} for a marginally higher gravity with $Bo=0.01$, $m=0.1092$, $\delta=5.19 \times 10^{-7}$ and $\epsilon=6.01 \times 10^{-8}$.

The various curves in the $\omega$ vs q plots in figures \ref{omega_q_1} and \ref{omega_q_2} are for different film thicknesses (non-dimensional). A DCM film of thickness $h=1.0$ signifies a film thickness of $2.35$mm. For a certain film thickness for an evaporating liquid film, the incidence of long wave instabilities decreases as the film thickness increases. In other words, thicker films are more resistant to long wave instabilities than thinner films for the corresponding liquid. As a film slowly evaporates, based on the gravity and thermocapillarity condition long wave instabilities may or may not manifest.

\begin{figure}
   \centering
   \includegraphics[width=\linewidth]{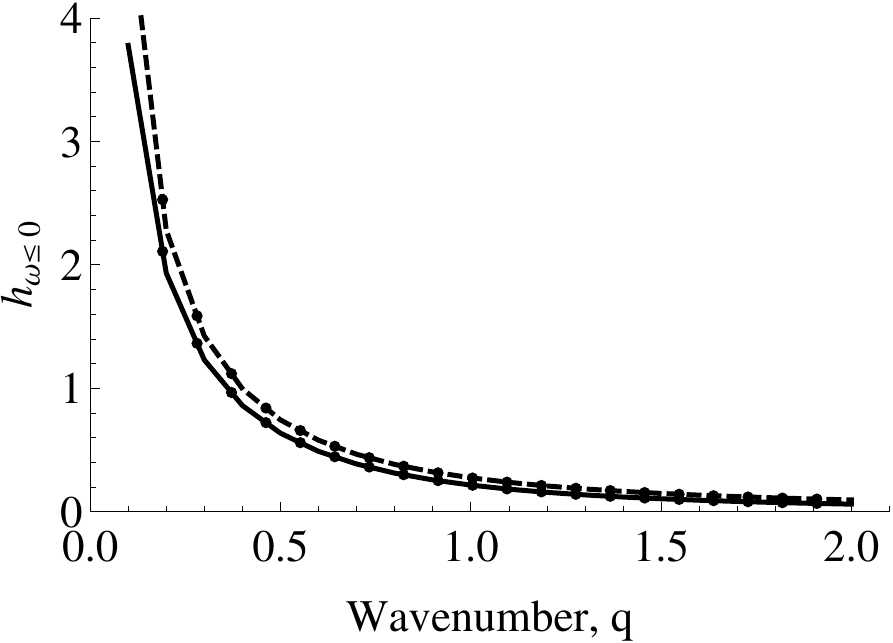}
   \caption{Film thickness is plotted for $\omega=0$ (solid) and $\omega=-0.0004$ (dashed) for different perturbation wavenumbers, q.This plot allows for the identification of the thickness of DCM film that is stable to long wave modes.}
   \label{h_vs_q}
  \end{figure}

From this observation, it is possible to back-calculate a film thickness for zero gravity evaporation for which the film is resistant to long wave instabilities. In figure \ref{h_vs_q}, DCM film thicknesses stable to long wave modes and those film thicknesses that represent zero growth rate are shown. Stabilization of long wave modes ($\omega \leq 0$) occurs for thicker films. As the film grows thinner by evaporation, a stable DCM configuration can be rendered unstable and shows the existence of long wave modes leading to thermocapillary driven rupture.

\section{Conclusions}

The evolution equation describing long wavelength instabilities affecting thin evaporating liquid films is used to capture film dynamics of an evaporating dichloromethane liquid film in zero gravity and in Earth's gravitational field. An expression for the fastest growing wavelength is derived. This expression allowed for zero gravity evaporation to be explored. The evolution equation and the expression for the fastest growing wavelength are validated against literature.

A $2.35$ mm thick dichloromethane liquid film evaporating in zero gravity shows interesting film dynamics when subject to a variety of initial condition and domain size combinations.

\begin{enumerate}
\item A DCM film evaporating in Earth's gravity is resistant to long wave type instabilities.
\item An evaporating DCM film in zero gravity confined to a square domain when perturbed with a smooth initial condition or a uniform random perturbation shows the appearance of the fastest growing wavelength at rupture.
\item When the domain shape is changed from a square to rectangle with length$=2.238\lambda_\text{max}$ and width$=1.641\lambda_\text{max}$, the fastest growing wavelength appears as the dominant mode at rupture. A directional preference is seen in the DFT plots.
\end{enumerate}

Zero gravity evaporation allows for evaporating DCM liquid films to be destabilized via long wave modes while a DCM film of the same thickness is robust and resilient to long wave type instabilities under the influence of Earth's gravitational field.

The growth rate of long wave instabilities is derived from the linear stability analysis is used as a metric to calculate the film thickness which provides the DCM film resistance to long wave type instabilities. Various growth rate curves are calculated for different film thicknesses and it is observed that when a DCM film is evaporating in zero gravity, as the film thickness decreases, it is more easily affected by long wave type instabilities.

\bibliographystyle{unsrtnat-clean}                

\bibliography{dnaneet_MASTER_03272013}

\clearpage

\section{Tables and Figures}

\begin{table}[h]
\centering
\begin{tabular}{|l|p{4cm}|p{4cm}|}
\hline
 Parameter & Definition & Physics described\\
\hline
$S$ & $\sigma h_0/\rho \nu^2$ & Surface tension \\
$G$ & $g h_0^3/\nu^2$  & Gravity\\
$M$ & $\sigma_T \Delta T h_0/\rho\nu\kappa$  & Thermocapillarity\\
Pr & $\nu/\kappa$ &  Momentum vs thermal diffusion\\
$E$ & $k \Delta T/\rho\nu L$ & Evaporation\\
$K$ & $\frac{k T_{sat}^{3/2}}{\alpha h_0 \rho^v L^2} \left(\frac{2 \pi R_g}{M_w}\right)^{1/2}$ & Non equilibrium at interface \\
$Ra$ & $g \beta \Delta T h_0^3/\nu\kappa$ & Buoyancy \\
$\sigma$ & Surface tension & \\
$h_0$ & Mean film thickness & \\
$T$ & Temperature & \\
$\rho$ & Mass density & \\
$\nu$ & Kinematic viscosity & \\
$\kappa$ & Thermal diffusivity & \\
$L$ & Latent heat & \\
$\alpha$ & Accommodation coefficient & \\
$R_g$ & Universal gas constant & \\
$M_w$ & Molecular weight & \\
\hline
\end{tabular}
\caption{Film parameters in evolution equation \ref{evolution1}}
\label{tab:nomenclature}
\end{table}
\clearpage

\begin{table}[ht]
\label{tab1}
\centering
\begin{tabular}{|p{4cm}|p{3cm}|p{3cm}|}
\hline
Case & Literature \citep{Oron2000b,Krishnamoorthy1995a} & MTU/UW\\
\hline
$S=100, Ga=0.333, M=35.1$ & 93.78 & 93.78 (validation)\\
$S=100, Ga=0.0, M=35.1$ & -- & 92.322\\
$E=0.0001, Ga=0.0, M=35.1, K=1.0$ & -- & 79.1712 \\
\hline
\end{tabular}
\caption{Comparison of fastest growing wavelength, $\lambda_\text{max}$. $\text{Pr}=7.02$ as per Oron et al. \citep{Oron2000b}, Krishnamoorthy et al. \citep{Krishnamoorthy1995a}. The first row shows a comparison/validation of the fastest growing wavelength as derived by us with equation with that of \citet{Oron2000b}}
\end{table}

\begin{center}
\end{center}
\begin{table}[h]
\centering
\begin{tabular}{|l|l|l|}
\hline
 & DCM ($g=9.81 / g=0.0 m/s^2$)  & Math fluid (regular/zero g)\\
\hline

Ga & $4.11 \times 10^6$ / 0 & 0.333 / 0\\
S & $143 \times 10^3$  & 100\\
M & $122 \times 10^3$  & 35.1\\
Pr & $3.9$ &  7.02\\
$\epsilon$ & $6.01 \times 10^{-8}$ & 0\\
$\delta$ & $5.19 \times 10^{-7}$  & 0 \\
m & $0.109$ & $0.05$ \\
\hline
\end{tabular}
\caption{Comparison of non-dimensional numbers for dichloromethane with the ``mathematical fluid''}
\label{dcm_math_comparison}
\end{table}

\vfill

\end{document}